# Ultra-efficient frequency comb generation in AlGaAs-on-insulator microresonators

Lin Chang,[1,†] Weiqiang Xie,[1,†,*] Haowen Shu,[1,2,†] Qifan Yang,[3] Boqiang Shen,[3] Andreas Boes,[1,4] Jon D. Peters,[1] Warren Jin,[1] Songtao Liu,[1] Gregory Moille,[5] Su-Peng Yu,[6] Xingjun Wang,[2] Kartik Srinivasan,[5] Scott B. Papp,[6] Kerry Vahala,[3] and John E. Bowers[1,*]

*[1] Department of Electrical and Computer Engineering, University of California, Santa Barbara, CA 93106, USA*

*[2] State Key Laboratory of Advanced Optical Communications System and Networks, Peking University, Beijing, 100871, China*

*[3] T. J. Watson Laboratory of Applied Physics, California Institute of Technology, Pasadena, California 91125, USA*

*[4] School of Engineering, RMIT University, Melbourne, VIC 3000, Australia*

*[5] Microsystems and Nanotechnology Division, National Institute of Standards and Technology, Gaithersburg, MD 20899, USA*

*[6] Time and Frequency Division, National Institute of Standards and Technology, Boulder, CO 80305 USA*

**Corresponding author: weiqiangxie@ucsb.edu, bowers@ece.ucsb.edu*

*†All three authors contributed equally to this work*

**Recent advances in nonlinear optics have revolutionized the area of integrated photonics, providing on-chip solutions to a wide range of new applications. Currently, the state of the art integrated nonlinear photonic devices are mainly based on dielectric material platforms, such as $Si_3N_4$ and $SiO_2$. While semiconductor materials hold much higher nonlinear coefficients and convenience in active integration, they suffered in the past from high waveguide losses that prevented the realization of highly efficient nonlinear processes on-chip. Here we challenge this status quo and demonstrate an ultra-low loss AlGaAs-on-insulator (AlGaAsOI) platform with anomalous dispersion and quality (Q) factors beyond $1.5 \times 10^6$. Such a high quality factor, combined with the high nonlinear coefficient and the small mode volume, enabled us to demonstrate a record low Kerr frequency comb generation threshold of ~36 µW for a resonator with a 1 THz free spectral range (FSR), ~100 times lower compared to that in previous semiconductor platform. Combs with >250 nm broad span have been generated under a pump power lower than the threshold power of state of the art dielectric micro combs. A soliton-step transition has also been observed for the first time from an AlGaAs resonator. This work is an important step towards ultra-efficient semiconductor-based nonlinear photonics and will lead to fully integrated nonlinear photonic integrated circuits (PICs) in near future.**

The extensive research on integrated nonlinear photonics in the last few years, driven by the breakthrough of the microcomb and other on-chip nonlinear devices, has opened up many new opportunities for on-chip integrated photonics, ranging from spectroscopy to atomic clock applications [1–3]. The demand to construct efficient nonlinear devices has motivated the development of different material platforms in nonlinear photonics. A common endeavor of those efforts is the reduction of the waveguide propagation loss, which is essential to enable high Q cavities so as to enhance the build-up power in the resonators and therefore increase the efficiency of the nonlinear optical processes [4]. In this regard, silica on silicon resonators [5–7] have long been dominant offering Q factors as high as 1 billion [6]. These devices can access a wide range of nonlinear effects including microwave rate soliton microcombs [8]. However, over the last 5 years, there has been remarkable progress to significantly improve the Q factors of resonators in many other nonlinear integrated optical material platforms. One example is the $Si_3N_4$ platform, which delivers high performance in Kerr comb generation on chip [9–11]. The $Si_3N_4$ micro-resonators have enabled the generation of efficient frequency combs with repetition rates from microwave to THz frequencies [12] and improved Q factor of beyond 30 million [13,14]. Another material, which recently attracted attention, is $LiNbO_3$. It offers additional opportunities for integrated nonlinear devices, due to its strong second order nonlinearity and electro-optic efficiency [15]. The significant reduction of waveguide loss in the lithium niobate on insulator (LNOI) platform [16,17] enabled the demonstration of resonators with Q factors beyond 10 million [18]. Its combined third and second order nonlinearities have also allowed soliton microcomb formation with intrinsic second-harmonic generation [19].

However, despite the significant progress, there are still many challenges in the current technologies of integrated nonlinear photonics. One problem is efficiency as measured by threshold or turn-on power. Focusing on Kerr-effect microcombs, all of the dielectrics noted above feature weak nonlinear Kerr indices ($n_2$ is usually in the $10^{-19}$ $m^2W^{-1}$ range or less). To reduce turn-on power and further the broad comb operation power to levels accessible by integrated lasers (tens of mW) [20], there are primarily two approaches in use. First is to boost optical Q factor, which provides a quadratic benefit to lowering the Kerr-effect parametric threshold [21]. However, to obtain high quality factors, the material absorption and waveguide scattering need to be carefully suppressed during microfabrication. A second approach is to reduce the mode volume of the resonator in order to decrease the required stored optical energy at threshold. Here, simply reducing the diameter of the resonator is not always an option as certain applications require large resonator diameters so as to obtain electronic-rate microcombs. Therefore, it is frequently necessary to turn to reduction of the modal cross sectional area to improve efficiency [14]. And while this approach has been particularly fruitful in silicon nitride microcombs, it does increase the challenge and complexity of maintaining high optical Q factor in the highly confined waveguides. As a result, it is common to have a large number of additional fabrication steps for high quality resonators [12,14]. These include chemical-mechanical polishing (CMP) and high temperature anneals. These additional steps can complicate the fabrication process, increase the cost, and sacrifice the yield and precise control of geometry, all of which add difficulties for high volume, low cost production in industry. A final and significant point is that the above dielectric nonlinear microcavities also encounter difficulties when they are integrated with active components due to incompatibilities in both material, design and fabrication, hindering the realization of fully integrated nonlinear photonic circuits.

One way to address these problems is to use III-V semiconductor materials. III-Vs feature extremely large nonlinear coefficients that are typically orders-of-magnitude larger than those of the dielectric materials noted above [22]. Also, their refractive indices are higher, which can be used to achieve a high index contrast and therefore a small modal cross sectional area (high intensity). These two factors are very attractive as they reduce the turn-on power of nonlinear optical processes in semiconductor-based nonlinear optical platforms. Moreover, III-V materials feature both a second and a third order nonlinear coefficient, both of which are required in frequency comb systems [23]. Furthermore, the wide usage of semiconductors in integrated photonics [20], as passive and active (laser and detector) elements, provides the potential for the integration of nonlinear devices into PICs.

However, most of the III-V semiconductor photonic platforms suffer from high waveguide losses, usually in the order of several dB/cm [24–26], which correspond to quality factors of ~$10^5$ or lower. The lack of sufficiently high Q in semiconductor platforms has limited the ability to harness their attractive material properties for nonlinear optics applications like microcombs. Recent demonstrations of whispering gallery mode or partially etched waveguide cavities in (Al)GaAs [27,28] achieved quality factor in the millions. This indicates that the current loss of integrated III-V waveguides can potentially be dramatically reduced. However, significant challenges remain to obtain such high Q factors within a waveguide that is both suitable for nonlinear applications such as Kerr microcombs or second harmonic generation (SHG) while also being compatible with integration. For example, fully dry etched surfaces are required in these structures as opposed to partially etched waveguides.

In this work, we make a key step in this direction by demonstrating compact micro-ring resonators in the AlGaAsOI platform with an intrinsic quality factor of beyond $1.5 \times 10^6$. The waveguides are fully etched with sub-micron dimensions and exhibit anomalous dispersion at a wavelength of ~1.55 µm. Their high quality factor, high Kerr nonlinear coefficients and compact mode volume, enabled us to demonstrate ultra-efficient frequency comb generation. For a 1 THz comb, the threshold power is only ~36 µW, which is a 100 times reduction relative to previous semiconductor resonators and 10 times lower compared to the state-of-the-art results from integrated dielectric micro-resonators. With 300 µW pump power we were able to generate a frequency comb that covers a spectral range of ~250 nm. A step transition of the comb state during laser scanning, which is the proof of soliton existence, has also be observed. Furthermore, the waveguide platform features a much simpler fabrication procedures compared to previous high-Q nonlinear platforms, and thus is suitable for high volume, low cost production. This demonstration paves the way to ultra-efficient nonlinear photonics of semiconductors, and provides a valuable solution to nonlinear PICs in the near future.

The nonlinear optical platform used in this work is AlGaAsOI. One key reason for selecting AlGaAs as a nonlinear material is its relatively large bandgap compared to other commonly used semiconductors in photonics, such as Si (1.1 eV (1127 nm)) or InP (1.34 eV (925 nm)). By changing the Al mole fraction, the bandgap of $Al_xGa_{1-x}As$ varies from 1.42 eV (872 nm) to 2.16 eV (574 nm) [29], which can be harnessed to avoid two photon absorption (TPA) at the two most important telecom bands (1310 nm and 1550 nm). In this work, we chose $x$ to be 0.2 for operating the comb at C band wavelengths. Higher Al portion can be used when targeting shorter pump wavelengths. Furthermore, AlGaAs has a very high nonlinear optical coefficient $n_2 = 2.6\times10^{-17}$ $m^2W^{-1}$, which makes it a very attractive in terms of nonlinear efficiency.

A schematic drawing of the fully etched AlGaAsOI waveguide cross section is shown in Fig. 1(a). One essential requirement for Kerr comb generation is that the waveguide should have anomalous group velocity dispersion (GVD) [3] at the pump wavelength. Our simulations indicated that, the AlGaAsOI waveguide needs to have sub-micron dimensions in order to obtain such dispersion at the two widely used telecom bands (O band and C band). This is due to the strong material dispersion of $Al_{0.2}Ga_{0.8}As$, which needs to be compensated by dispersion engineering tailoring the waveguide geometry. In this work, the AlGaAs layer thickness is set to be 400 nm, at which the calculated GVD is anomalous at C band wavelength for waveguides with several different widths, as shown in Fig. 1(c). The simulated mode distribution of one example waveguide (400 nm × 700 nm) is plotted in Fig. 1 (b). Compared to the commonly used $Si_3N_4$ waveguides for comb generation, the mode volume is reduced by a factor of ~4, which not only enhances the photon intensity, but also enables more compact devices.

The fabrication of the AlGaAsOI platform is based on heterogeneous wafer bonding technology, similar to the previous process of GaAsOI devices used for second harmonic generation [30], and is discussed in detail in the Method part.

The propagation loss of the waveguides needs to be low in order to achieve high Q resonators. This requires the reduction of the sidewall roughness, which plays a major role in limiting the loss of the AlGaAsOI platform. There are two reasons for this: first, the mode size of the waveguide is small, which leads to a significant interaction of the mode with the waveguide sidewall; second, the strong index contrast between AlGaAs and $SiO_2$ causes an increased scattering loss, which scales with $(n_{core}^2 - n_{clad}^2)$[31]. As a result, it is critical to lower the sidewall and surface roughness in all high index contrast platforms in order to obtain low propagation loss.

In this work we reduced the scattering loss mainly by two means. The first one is the implementation of a reflow process of the patterned photoresist after the lithography process. Fig. 2 (a) shows the SEM top-view images of the patterned $SiO_2$ hard mask with and without reflow of photoresist, both of which are exposed using an ASML 248 nm DUV stepper. It can be seen that without reflow, a significant amount of roughness is visible at the edge of the $SiO_2$ hard mask, which would be transferred to the sidewall of waveguide after AlGaAs etch and would act as a strong optical scattering source. When a resist reflow step is applied after the lithography process, the resist boundary is smoothed out and the roughness is hardly visible in the SEM image. The change of resist shape caused by reflow can be pre-calibrated and taken into account in mask design, which enables dimensional control of the waveguide width with nanometer scale.

The second means to reduce the scattering loss is an optimized dry etch process. Here we applied an Inductively Coupled Plasma Etching (ICP) etch for both the hard mask and the AlGaAs. The gases we used are $CHF_3/CF_4/O_2$ for etching the $SiO_2$ and $Cl_2/N_2$ for etching the AlGaAs. The well-developed etching recipes resulted in smooth etching profile as shown in Fig. 2 (b). The SEM picture of the cross section of the waveguide (Fig. 2 (c)) shows that the sidewall angle is very close to 90 degrees, which is beneficial for accurate geometry control.

Besides the scattering, another loss contributor for semiconductor waveguides is the absorption caused by defect states at material surfaces. Recent research shows that this factor starts to become dominate for resonators operating in the high Q region [27]. A way to reduce loss originating from this source is to apply a surface passivation treatment to the waveguide surface, which can eliminate the intra-band states. In this work, a 5-nm thick $Al_2O_3$ layer was deposited by Atomic Layer Deposition (ALD), which fully surrounds the waveguide core and passivates the AlGaAs surface. Other methods to reduce the surface dissipation, such as wet nitridation, have also been discussed in the literature [27] and may further improve the passivation.

As indicated above, to generate a frequency comb in ring resonators, the dispersion of the waveguide plays a critical role. To characterize the GVD of the waveguides we fabricated a ring resonator with 100 µm radius and a free spectral range (FSR) of 118 GHz. For this resonator we measure the resonance frequency of a mode family as a function of relative mode number $\mu$, relative to a reference resonance at $\omega_0$, which is around 1550 nm. The resonance frequency $\omega_\mu$ of the modes can be expanded in Taylor series as:

$$\omega_\mu = \omega_0 + \mu D_1 + \frac{1}{2}\mu^2 D_2 + \frac{1}{6}\mu^3 D_3 + \cdots \quad (1)$$

Where $D_1/2\pi$ refers to the FSR around $\omega_0$ and $D_2$ is related to the GVD $\beta_2$ by $D_2 = -\frac{c}{n} D_1^2 \beta_2$. Fig. 3 (a) shows the measured relative mode frequencies $D_{int} \equiv \omega_\mu - \omega_0 - \mu D_1$. By fitting the data, second order dispersion $D_2/2\pi$ is extracted to be 10.8 MHz. This confirms that the resonator is operated in the anomalous dispersion regime.

The quality factor of the resonators is estimated based on the transmission spectrum of the resonances. Fig. 3 (b) shows a resonance at a wavelength of ~1519 nm for a 1THz ring resonator, which has waveguide width of 700 nm and radius of 12 µm. The intrinsic quality factor is extracted to be ~$1.53 \times 10^6$, which correspond to a propagation loss around 0.4 dB/cm. The Q factor is one order of magnitude higher compared to that of previous AlGaAsOI resonators for comb generation with a similar radius [32]. Considering the compact mode size (~0.28 $\mu m^2$) and small radius of the ring, it is reasonable to believe that this platform has waveguide losses that are comparable to the state of the art fully etched silicon on insulator (SOI) or even many commonly used dielectric waveguides.

Fig. 3 (c) presents the transmission spectrum of a resonance, which shows a splitting of the mode. This phenomenon is usually observed in high-Q resonators [33], which are sensitive to small imperfections or scatters at waveguide surfaces. This indicates that the quality factors are still limited by the scattering loss and can be further reduced by optimizing the fabrication process.

Next we investigate the nonlinear optical efficiency of the waveguides to illustrate the advantage of combining high nonlinear coefficients, high index contrast and high quality factors simultaneously on this AlGaAsOI platform. Fig. 4(a) shows the parametric oscillation spectrum of such a 1 THz resonator operated at 36 µW, where one can observe the onset of frequency comb generation. This indicates that we were able to generate efficient Kerr combs by pumping the resonators at C-band wavelengths. The threshold power is ~100 times lower compared to previous AlGaAsOI resonators with similar FSR [32] and is consistent with the fact that the Q factor has a 10 times improvement.

Besides the significant reduction of the threshold power, the high $n_2$ of AlGaAs highlights its advantages even more dramatically in terms of the efficiency of comb broadening above the threshold. It can be seen in Fig. 4(b) that for the same 1 THz resonator discussed above, the generated comb lines cover a >250 nm wide spectral range under a power of only 300 µW. For another comb with 450 GHz FSR, 250 µW pump power enables a comb span over 200 nm with > 50 comb lines. It is worth to point out that these pump powers are still lower than the record threshold power reported for $Si_3N_4$ microcomb [14]. Under such quality factor, only a few mW pump power should be sufficient to generate an octave span THz comb with a double dispersive wave if proper dispersion engineering is applied, which is essential for *f-2f* self-referencing and frequency synthesis applications [1].

A comparison of different nonlinear material platforms for frequency comb generation is shown in Table. 1. The demonstration in this work, to the best of our knowledge, delivers the lowest threshold power for comb generation so far. It needs to be noted that for the $Si_3N_4$ [14] and silica [7,21,34] platforms with the highest Q, considering the scale factor, they may have similar level threshold for a THz comb assuming the Q is constant with those of the big resonators. This however is not the case, because for small ring radius their waveguide loss usually increases significantly either due to enhanced bending or scattering loss [21,35]. On the other hand, it also indicates that the Q factor of AlGaAsOI resonator can be even higher for larger rings.

The key advantage offered by using AlGaAs is its high Kerr nonlinear coefficient ($n_2$ = 2.6×10$^{-13}$ cm$^2$W$^{-1}$), which is two orders of magnitude higher than that of $Si_3N_4$ ($n_2$ = 2.5×10$^{-15}$ cm$^2$W$^{-1}$). Considering the expression for the parametric oscillation threshold power ($P_{th}$) of a comb [21,36]:

$$P_{th} \approx 1.54(\frac{\pi}{2})\frac{1}{\eta}\frac{n}{n_2}\frac{\omega}{D_1}\frac{A}{Q_T^2} \quad (2)$$

where $\eta = \kappa_e/\kappa$ is the coupling factor ($\kappa$ and $\kappa_e$ are total and cavity related decay rate), $n_2$ $(n)$ refers to the Kerr nonlinear index (refractive index), $A$ is the mode area and $Q_T$ refers to the total quality factor of the resonator. To highlight the importance of the highly nonlinear optical material and small mode volume, we consider an example comparing the influence of two different material systems on the threshold of comb power. For this exercise we assume that the resonators in the different materials are operated at the same frequency with the same FSR, and the same coupling parameter. In this case, only the nonlinear coefficient, mode area, quality factor and the refractive index of the waveguide determine the threshold power. For AlGaAsOI waveguides ($n$ = 2.9, $A$ = 0.28 μm$^2$), one can obtain that for the same Q factor, the threshold power for AlGaAsOI resonator will be ~270 times lower compared to that of $Si_3N_4$ standard micro-resonator. In another words, in order to achieve the same threshold power, the quality factor required for AlGaAsOI resonator is 16 times lower than the state of the art $Si_3N_4$ technology, and it moreover relaxes the strict fabrication requirements to achieve low loss waveguides significantly.

Another advantage, offered by AlGaAs or other epitaxial grown materials is precise thickness control. Most of the dielectric thin films, either prepared by direct deposition or smart-cut bonding technology, usually require chemical-mechanical polishing (CMP) to reduce their surface roughness and thereby boost optical Q factor. This process removes a significant amount of material and introduces a non-uniformity to the film thickness on the order of tens of nanometer over the wafer scale. As a result, such ultra-high-Q resonators can suffer from significant waveguide geometry variation, which is problematic in many applications that rely upon accurate control of dimensions. For example, the detection of carrier envelope offset frequency of microcombs for self-referencing or the efficient spectral translation between visible and infrared light require precise dimensional control of the waveguide cross section [37]. Compared to dielectrics, the epitaxial growth for semiconductors enables atomic scale accuracy of material thickness with high uniformity, providing a good solution to dimension control in nonlinear photonics.

One commonly used step in the fabrication of high Q resonators, but not critical in AlGaAsOI process, is a high temperature (>1000°C) anneal, which reduces the O-H and N-H bonds in deposited $Si_3N_4$ and $SiO_2$ layers, as such bonds cause absorption losses. However, this step is not compatible with the fabrication of standard photonic foundries, especially the complementary metal–oxide–semiconductor (CMOS) process. For AlGaAsOI waveguide, more than 90 % of the mode is confined inside the waveguide core so that the cladding contributes very little to the total loss. Moreover, the cladding loss is negligible given the fact that the quality factor in this case is greatly relieved.

To further utilize such efficient frequency comb generation in AlGaAsOI platform, the next essential step is to access soliton formation, which enables stable and low noise states for comb operations [38]. Usually fast frequency scanning of the pump laser is required to kick off the soliton formations in integrated micro-resonators, due to the fast thermal timescales. When the pump frequency is swept across the resonances, the transition from modulation instability (MI) state to soliton regime leads to a large drop in comb power and therefore a large temperature change, shifting the cavities resonances and making the soliton state unstable and hard to get access to. As a result, the frequency scanning speed of pump laser needs to be fast for resonators with high absorption and strong thermo-optic effects. For AlGaAsOI resonators, this problem becomes even more challenging than dielectrics, because AlGaAs' thermo-optic coefficient ($2.3\times 10^{-4}$ $K^{-1}$) is one order of magnitude higher compared to that of $Si_3N_4$ ($2.4\times 10^{-5}$ $K^{-1}$) or Silica ($0.8\times 10^{-5}$ $K^{-1}$). However, on the other hand, thanks to highly efficient process in this platform, the pump power required for soliton generation is much lower compared to the power in previous experiments, which relieves thermal effect inside the cavities and therefore the requirement of scanning speed of laser. Here we observed a step transition from MI to stable comb state during laser scanning, as shown in Fig. 5, which indicates the existence of soliton. The scan speed is only 5 nm/s and pump power is 2.5 dBm in the bus waveguide for a 450 GHz comb. At this point, due to the relatively large thermal triangle of the resonance, it is challenging to stop the laser frequency at the soliton step and extract the spectrum. Further improvement in the Q will reduce the power requirement for soliton generation and facilitate the investigation of soliton formation in this platform.

Instead of the traditional method, recently several approaches demonstrated provide other solutions to overcome the thermal problem and generate low-noise comb state, such as using an auxiliary laser for providing temperature compensation [39], or applying the dispersion engineering in normal GVD region to eliminate the influence of thermal effect [40]. Further investigation along those directions will be executed to explore the potential of the ultra-efficient combs in this platform.

In addition to the microcomb discussed above, the AlGaAsOI platform with ultra-low loss also provides solutions to realize many other nonlinear processes with ultra-high efficiencies on chip. One example is second order frequency conversion [27], which is of importance for various applications such as $f$-$2f$ self-referencing. Previously, second harmonic generation (SHG) with record high conversion efficiency based on (Al)GaAsOI waveguides and resonators has been demonstrated based on GaAsOI waveguide [30][41]. Assuming that the 6 times higher quality factor as demonstrated in this work is applied to the SHG resonators, the efficiency can be improved by a factor of $(Q_{Pump})^2 Q_{SHG}$, which is more than 200 and in this case is $10^8$ $\%W^{-1}$ level. Since AlGaAsOI supports both of the most efficient second and third order nonlinear effects in integrated photonics, it holds the potential to simultaneously incorporate self-referencing using SHG together with octave-spanning Kerr comb generation. It should also be noted that in quantum optics, the efficient $\chi^{(2)}$ and $\chi^{(3)}$ processes are important for several processes, e.g. the spontaneous parametric down-conversion [42] and spontaneous four wave mixing [37]. Moreover, other nonlinear properties of AlGaAs, such as the photoelastic and piezoelectric effects [43], can be harnessed by utilizing the high Q cavity in optomechanics.

Besides the significance in nonlinear and quantum photonics, this work will also have far-reaching impact to traditional PICs', both in academia and industry. The AlGaAsOI waveguide structure now enables similar performance compared to a fully etched SOI waveguide [44], but with much lower nonlinear loss. This makes such a platform suitable for high power and narrow linewidth lasers, optical buffer devices and photonic sensors [45]. Furthermore, compared to SOI, the epi of (Al)GaAs is compatible with the III-V laser gain medium, and therefore can dramatically simplify the current integration technology of heterogeneous III-V on Si, without sacrificing device performance.

In conclusion, we demonstrated a low loss AlGaAsOI platform and applied it to create microcavities with a $Q$ factor beyond $1.5 \times 10^6$. These high-Q devices in combination with the high nonlinearity of the AlGaAsOI waveguides enabled ultra-efficient frequency combs. A record low threshold power around 36 µW for 1 THz comb was achieved. Moreover, pumping the resonator with 300 µW power leads to an efficient comb broadening to a span over 250 nm. The observation of a soliton step in AlGaAsOI platform has also been reported. This demonstration opens up many opportunities for ultra-efficient nonlinear photonics in this platform. Furthermore, it paves the way for fully integrated nonlinear PICs in the future.

**Methods:**

**Fabrication.** In this work, the layer structure of the epi from top to bottom is: a [001] orientated 400-nm-thick $Al_{0.2}Ga_{0.8}As$ film on a 500-nm-thick $Al_{0.8}Ga_{0.2}As$ layer grown on a 500-µm-thick GaAs substrate. A 5-nm-thick $Al_2O_3$ layer was deposited on the top surface of the epi by Atomic Layer Deposition (ALD) for passivation. This chip was then bonded onto a Si wafer with 3-µm-thick thermal $SiO_2$ layer, after plasma activation. The thermal $SiO_2$ layer was patterned before the bonding by inductively coupled plasma (ICP) etching to form cross-shape vertical channels (VCs) with 50 µm spacing for gas release. The bonded piece was annealed at 100˚C for 24 hours under pressure to enhance the bonding strength. Afterwards, mechanical polishing was applied to lap the GaAs substrate thickness down to 70 µm. The remaining GaAs substrate was removed by wet etching with $H_2O_2$:$NH_4OH$ (30:1) and the $Al_{0.8}Ga_{0.2}As$ layer was removed by diluted hydrofluoric acid.

After substrate removal, $SiO_2$ hardmask was deposited on the $Al_{0.2}Ga_{0.8}As$ thin-film, which was patterned by using Deep Ultraviolet (DUV) lithography followed by a reflow proceduer of the photoresist. ICP etching with a $CHF_3/CF_4/O_2$ gas chemistry was then applied to etch the hard mask. Afterwards, an ICP etching step with a $Cl_2/N_2$ gas chemistry was applied to etch the $Al_{0.2}Ga_{0.8}As$ layer. Finally, another layer of $Al_2O_5$ was used to passivate the sidewall and the sample was then coated with $SiO_2$ to protect the waveguides.

**Frequency comb characterization.** For testing frequency combs, we used two kinds of ring resonators in this work with 1 THz and 450 GHz FSR, both of whose waveguide cross section geometry is 400 × 700 nm. The chip facets have an inverse taper structure, interfaced with lensed fiber. The coupling loss per facet is around ~5 dB. The light source used in the experiment is a tunable laser at C band (Keysight 81608A). When measuring the threshold power and broad comb operation, the pump laser is slowly tuned into the resonance (from red detune to blue) of resonator and the transmitted pump light and generated comb spectrum is recorded by an optical spectrum analyzer (OSA) (YOKOGAWA AQ6375). For the soliton generation experiment, the laser is automatically swept over the resonance by using laser's build-in scanning function. The output fiber is connected to a waveshaper (Finisar 4000A), splitting the pump light and the generated other comb lines. The two channels of light are received by two photodetectors (New Focus, Model 1811) respectively, which are monitored by an oscilloscope (Tektronix MSO64) for recording power traces.

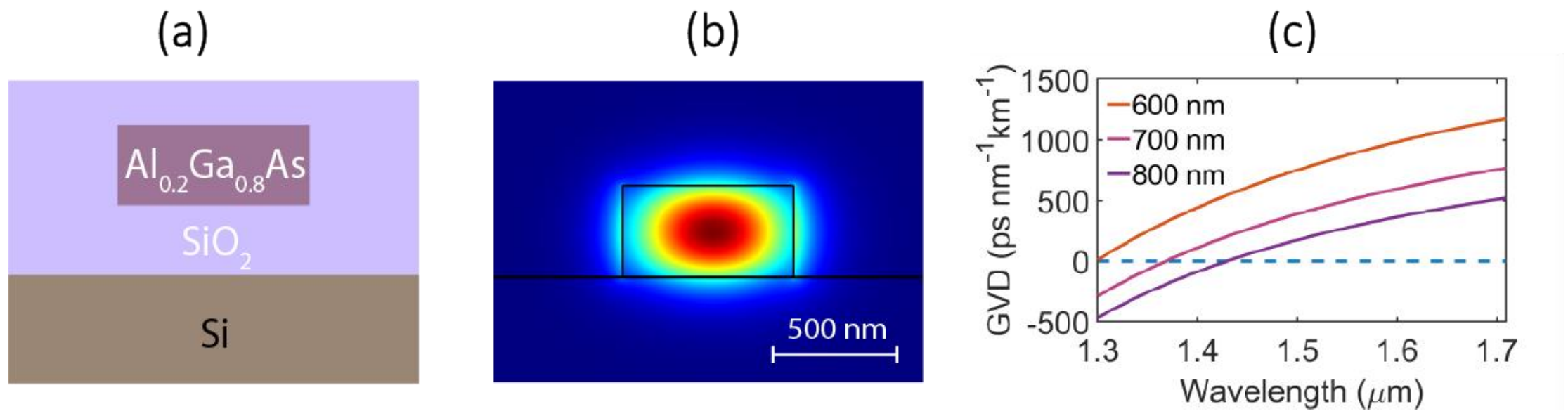

**Figure 1:** (a) Schematic drawing of the AlGaAsOI waveguide cross section; (b) simulated intensity distribution of the waveguide fundamental TE mode for comb generation; (c) simulated GVD of 400 nm thick AlGaAsOI waveguides with different widths.

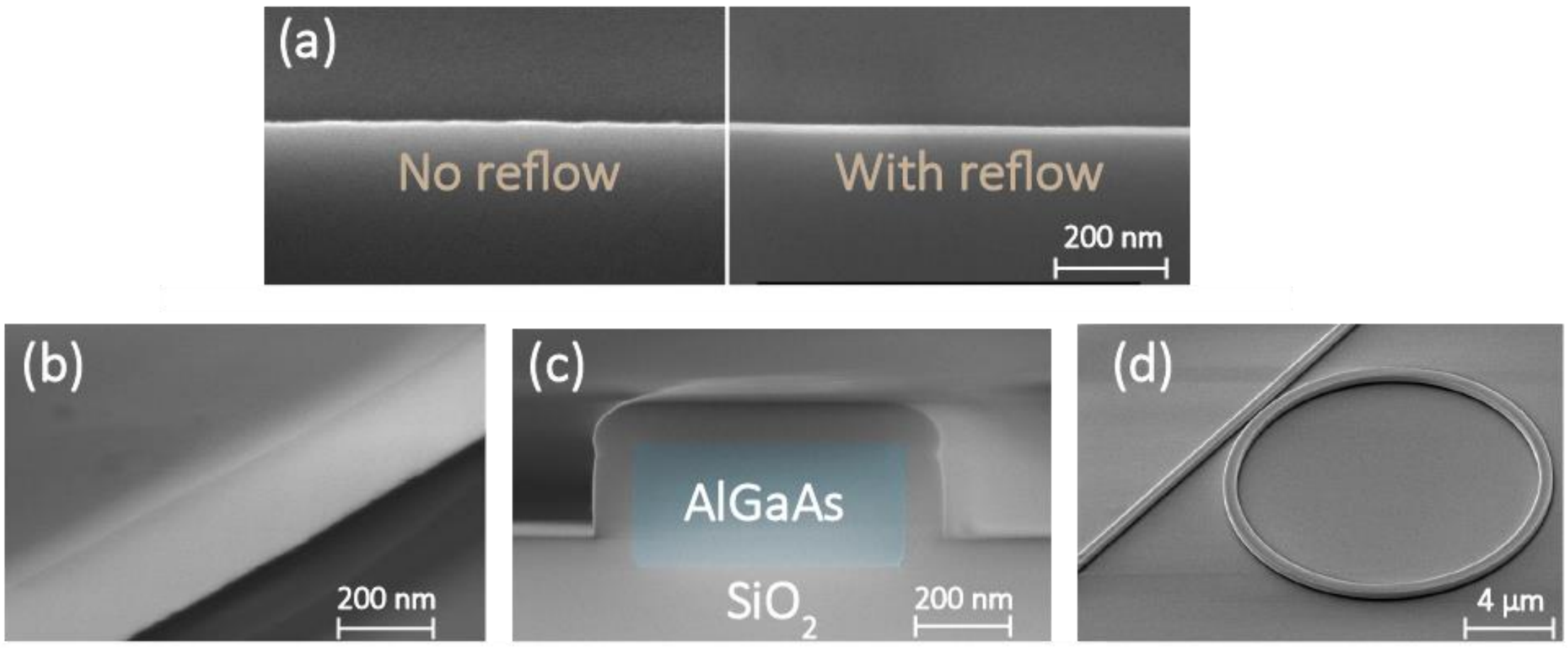

**Figure 2:** SEM images of (a) top view of the $SiO_2$ hard masks with and without reflow applied after the lithography process; (b) sidewall of the waveguides; (c) cross section of the waveguide after passivation and a thin layer of $SiO_2$ deposition; the AlGaAs core is highlighted with false color (blue); (d) a ring resonator.

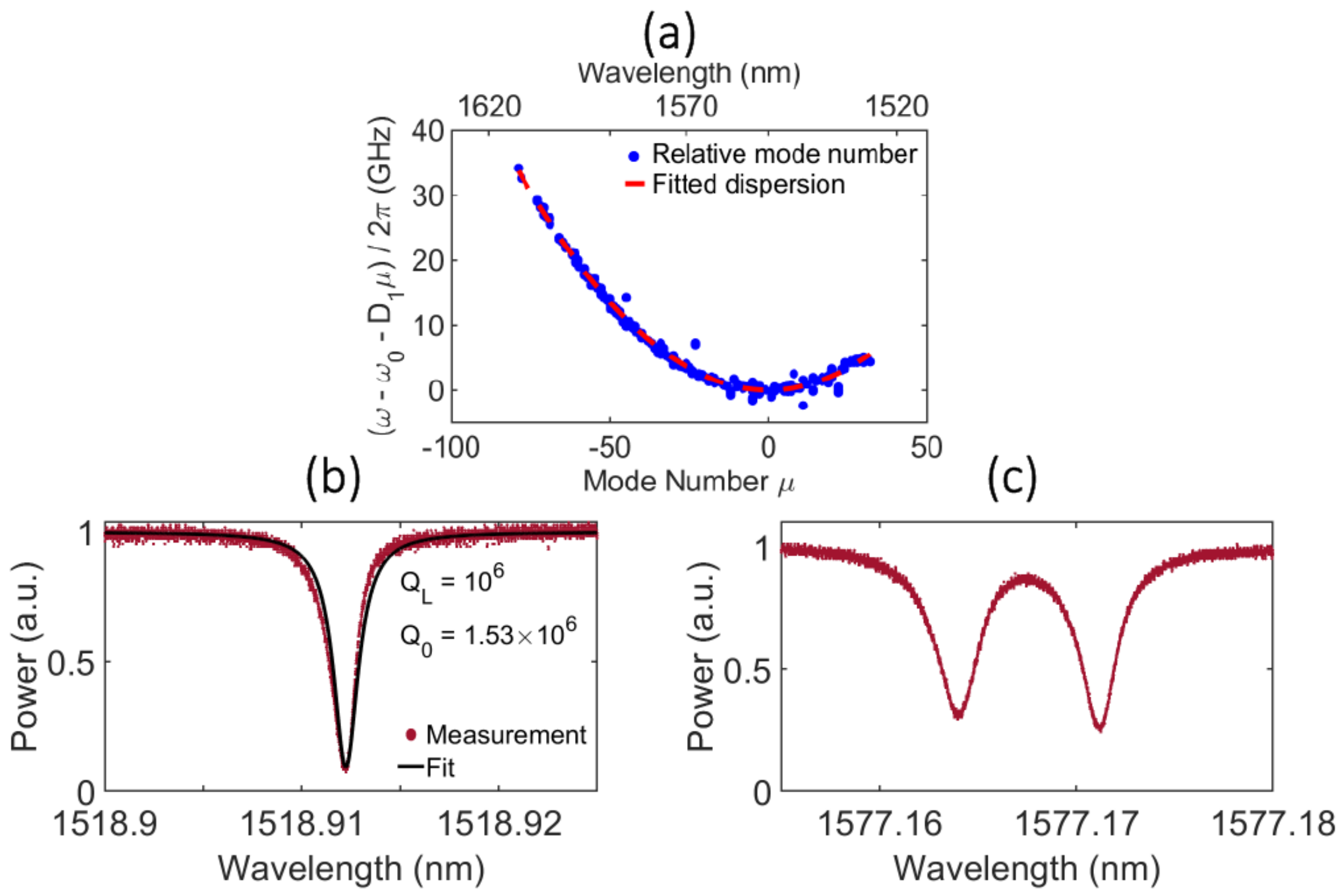

**Figure 3:**Experimental results: (a) Measured relative mode frequencies $D_{int}$ plotted verse μ; (b) Measured transmission spectrum of a resonance around 1518 nm (red dots) and fitting curve (black line); (c) Resonance with splitting due to backscattering (red dots).

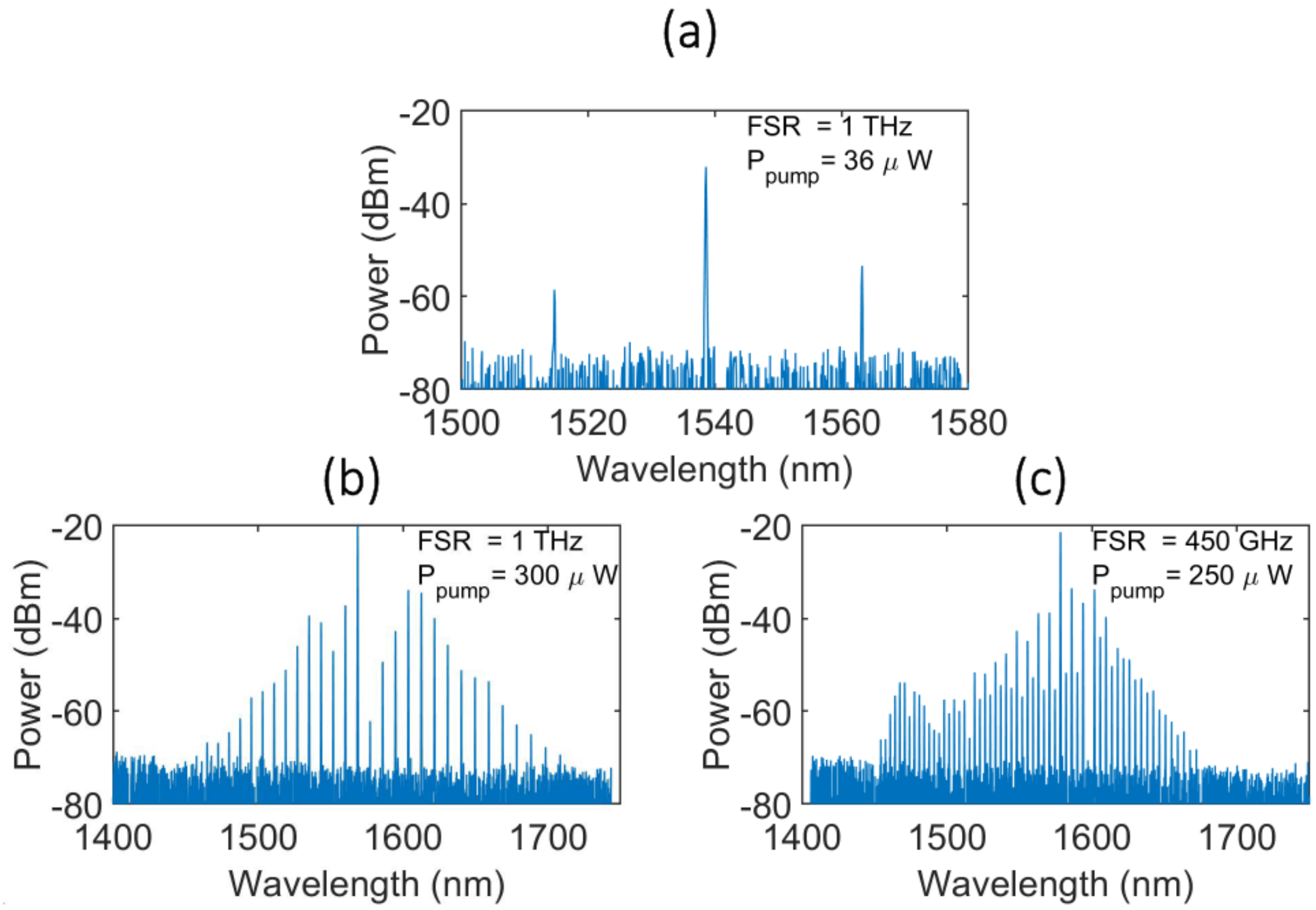

**Figure 4:** Frequency comb spectrum by a 1 THz resonator under pump power of (a) 36 μW and (b) 300 μW, and a 450 GHz resonator under pump power of (c) 250 μW.

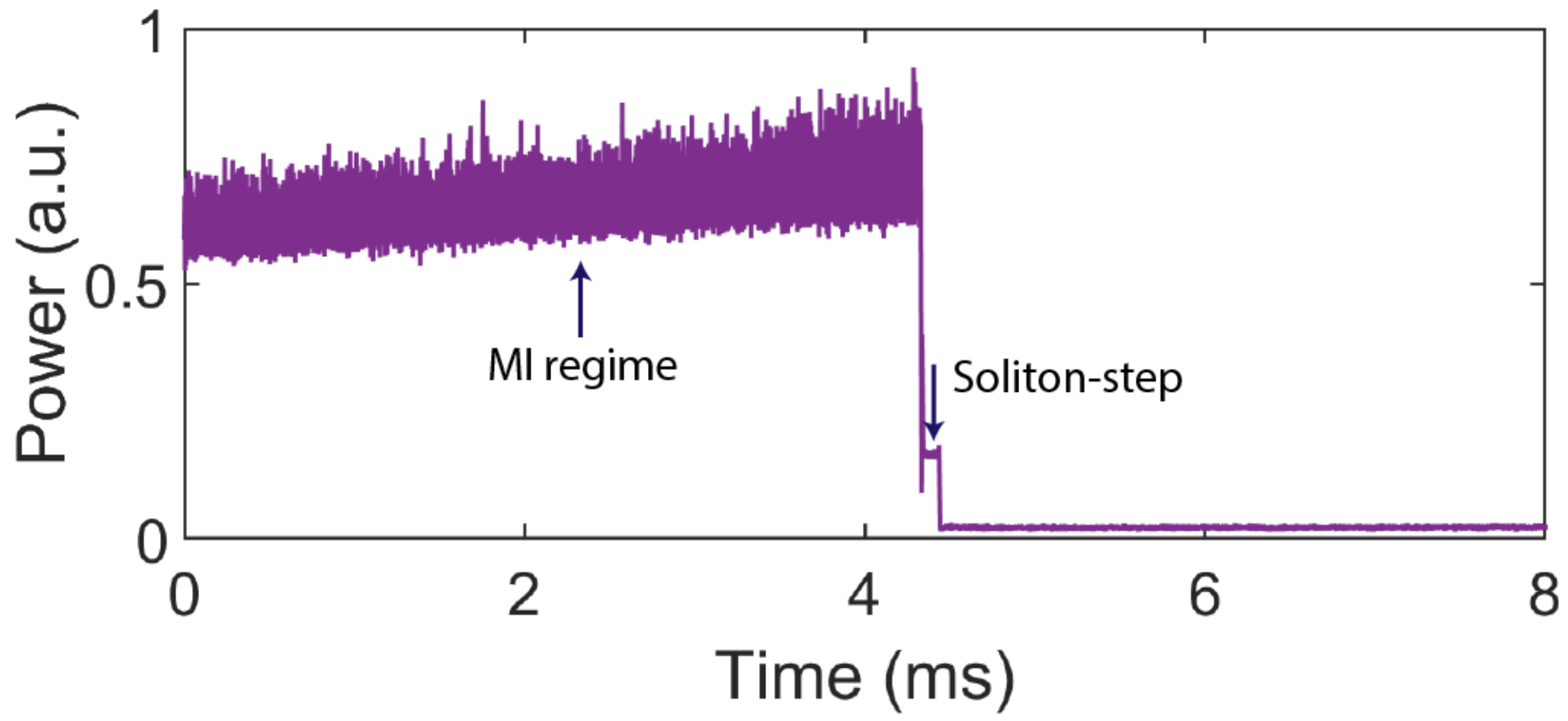

**Figure 5:** Response of comb power when laser is swept through the AlGaAs resonance.

The step-like trace indicates a transition to the soliton state.

| Material | Refractive index | $n_2$ ($m^2W^{-1}$) | Mode area ($\mu m^2$) | Highest Q for comb generation | Lowest threshold power (mW)(FSR) |
|---|---|---|---|---|---|
| Silica [34] (Wedge disk) | 1.45 | $3\times10^{-20}$ | ~60 | $6.7 \times 10^8$ | 1.2 (9.3 GHz) |
| Silica [21] (Microtoroid) | | | ~10 | $1.2 \times 10^8$ | 0.17 (1 THz) |
| $Si_3N_4$ [14] | 2.0 | $2.5\times10^{-19}$ | ~1.5 | $3.7 \times 10^7$ | 0.33 (200 GHz) |
| $LiNbO_3$ [19] | 2.21 | $1.8\times10^{-19}$ | ~1 | $2.2 \times 10^6$ | 4.2 (199.7 GHz) |
| $Ta_2O_5$ [46] | 2.05 | $6.2\times10^{-19}$ | ~1.5 | $3.2 \times 10^6$ | 10 (500 GHz) |
| Hydex [47] | 1.7 | $1.15\times10^{-19}$ | ~2 | $1 \times 10^6$ | 50 (200 GHz) |
| Si [48] | 3.47 | $5\times10^{-18}$ | ~2 | $5.9 \times 10^5$ | 3.1 (127 GHz) |
| GaP [49] | 3.05 | $7.8\times10^{-19}$ | ~0.15 | $3 \times 10^5$ | 10 (500 GHz) |
| AlN [50] | 2.12 | $2.3\times10^{-19}$ | ~1 | $9.3 \times 10^5$ | NA |
| AlGaAs (this work) | 3.3 | $2.6\times10^{-17}$ | ~0.28 | $1.5 \times 10^6$ | 0.036 (1 THz) |

**Table. 1**. Performances of various nonlinear materials for microcomb generation

**Acknowledgements**

This work is supported by a DARPA MTO DODOS contract (HR0011-15-C-055). We thank Gordon Keeler, Justin Norman, Mario Dumont, Garrett Cole, Junqiu Liu, Eric Stanton, Richard Mirin for fruitful discussions.

**Author contributions**

L.C., W.X., H.S. and J.B. conceived the experiment, H.S., L.C., G.M. and S.Y. did the simulation and design for the devices. W.X., L.C. and J.P. did the fabrication. W. J. and S.L. assisted in the fabrication. L.C., W.X., H.S., Q.Y. and B.S. performed the experiment. L.C., W.X., H.S., Q.Y, B.S., A.B., K.S. and S.P. analyzed the data. All the authors contributed in writing the manuscript. X.W., K.S., S.P., K.V. and J.B. supervised this project.

**Additional information**

The authors declare no competing financial interests.

This research was developed with funding from the Defense Advanced Research Projects Agency (DARPA). The views, opinions and/or findings expressed are those of the author and should not be interpreted as representing the official views or policies of the Department of Defense or the U.S. Government.